\documentclass[%
 prl,preprint,groupedaddress,
 amsmath,amssymb,
 aps,
]{revtex4-1}

\usepackage{graphicx}
\usepackage{dcolumn}
\usepackage{bm}

\begin{document}


\title{Searching for Stoponium along with the Higgs boson}

\author{Vernon Barger}
 \email{barger@wisc.edu}
\affiliation{
Department of Physics, University of Wisconsin, Madison, WI 53706
}%

\author{Muneyuki Ishida}
 \altaffiliation[ ]{Department of Physics, University of Wisconsin-Madison. A visitor until March 2012.}
 \email{mishida@wisc.edu}
\affiliation{%
Department of Physics, School of Science and Engineering, Meisei University, Hino, Tokyo 191-8506, Japan
}%


\author{Wai.-Yee Keung}
 \email{keung@uic.edu}
\affiliation{
Department of Physics, University of Illinois, Chicago, IL 60680, USA
}%


\date{\today}

\begin{abstract}
Stoponium, a bound state of top squark and its antiparticle in a supersymmetric model, 
may be found in the ongoing Higgs searches at the LHC. 
Its $WW$ and $ZZ$ detection ratios relative to the Standard Model Higgs boson can be more than unity 
from $WW^*$ threshold to the two Higgs threshold.
The $\gamma\gamma$ channel is equally promising. A stoponium mass 135 to 150~GeV
is sererely constrained by the ATLAS and CMS experiments.   
\end{abstract}

\maketitle


Discovery of a Higgs boson $h^0$ is a top priority of LHC experiments, 
as is the search for the supersymmetry which is an attractive candidate for physics beyond the SM.
Here we make the observation 
that there is a possibility of finding supersymmetry in the LHC search for the Higgs boson. 

The stop $\tilde t_1$, a scalar superpartner of the top quark, is expected to have the lightest mass
of all the squarks\cite{Rudaz}, and may plausibly be lighter than the top quark.   
If the mass difference between $\tilde t_1$ and the LSP neutralino $\tilde N_1$ is small and the tree-level decays,
$\tilde t_1\rightarrow t\ \tilde N_1$ and $\tilde t_1\rightarrow b\ \tilde C_1$ (where $\tilde C_1$ is the
 lightest chargino), are not kinematically allowed, the $\tilde t_1$ becomes long-lived 
and $\tilde t_1^*\ \tilde t_1$ will form a bound state called stoponium. 
This expectation is supported by the result\cite{HK} that the partial width of 
$\tilde t_1\rightarrow c\ \tilde N_1$ occurring at loop level 
is negligibly small compared with the binding energy of the stoponium, a few GeV. 
 
The possibility of stoponium discovery at hadron colliders was considered long ago\cite{HMR,BK}.
Its production via $gg$-fusion and its decays are quite similar to the 
heavy quarkonium of the fourth generation quarks\cite{BGHK}.
The production amplitude is proportional to the wave function at the origin,
and so the $S$-wave $J^{PC}=0^{++}$ bound state, denoted here as $\tilde\sigma$,
is expected to have the largest production cross section.

Detection of stoponium is complementary to the detection of stop.
Actually current LHC data already give a strong limit on the light stop mass\cite{PRW}
by considering the possible decay channels of $\tilde t_L\rightarrow \tilde b_L W^*$
and $\tilde t_L\rightarrow t \tilde N_1$. For the parameter space where these decay modes are kinematically
forbidden and stop is long-lived, as required for stoponium existence, this constraint does not apply.
The stop decays to charm quark and LSP neutralino at loop-level, and this decay mode is notoriously
difficult to be identified, because of hadronic effects and the small phase space. 
In this parameter region, stoponium detection may be the best way to
search for supersymmetry. 

In this Letter we refine the calculation of the production and decays of the $\tilde\sigma$
appropriate to the LHC experiments at 7 TeV (LHC7) 
and consider the possibility of finding the $\tilde\sigma$ in the ongoing LHC Higgs searches. 
Related work can be found in ref.\cite{DN}, and more recently in ref.\cite{Martin}, at tree level, 
and the NLO radiative corrections are considered in refs.\cite{MY,YM}.
We demonstrate that $\tilde\sigma$ could be found in the SM Higgs search in 
the $\gamma\gamma$ and $W^*W^*$ channels at LHC7.
Stoponium can be distinguished from a Higgs boson by differences in decay branching fractions.\\

\begin{figure}[htb]
\begin{center}
\resizebox{0.7\textwidth}{!}{
  \includegraphics{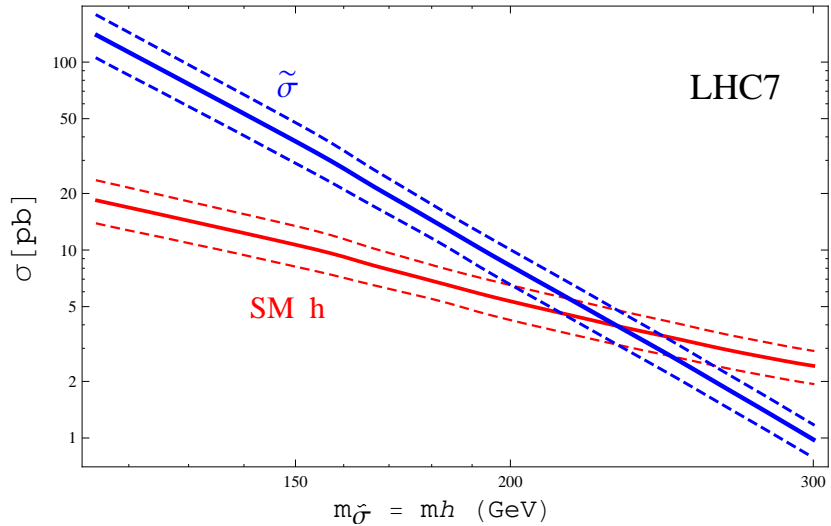}
}
\end{center}
\caption{The production cross section of $\tilde\sigma$[pb] from $gg$ fusion (solid blue), 
$\sigma(gg\rightarrow \tilde\sigma)$[pb],
compared with that of the SM Higgs with the same mass $m_{h^0}=m_{\tilde\sigma}$(solid red).
The overall theoretical uncertainties\cite{Djouadi} are denoted by dotted lines.
}
\label{fig1}
\end{figure}

\noindent\underline{\it Stoponium Production Cross Section}\ \ \ 
The production cross section of stoponium $\tilde\sigma$ in hadron colliders
is mainly via $gg$ fusion, similarly to the production of a Higgs boson $h^0$.
The cross sections are proportional to the respective partial decay widths to $gg$.
The production cross section of $h^0$ has been calculated in NNLO\cite{Djouadi},
and by using this result\cite{comment0} we can directly estimate the production cross section of $\tilde\sigma$ as
\begin{eqnarray}
\sigma(pp\rightarrow \tilde\sigma X) &=& \sigma(pp\rightarrow h^0 X)\times
\frac{\Gamma(\tilde\sigma\rightarrow gg)}{\Gamma(h^0\rightarrow gg)} \ .
\label{eq1}
\end{eqnarray}
By using the $\Gamma(\tilde\sigma\rightarrow gg)$ partial width given later and 
$\Gamma(h^0\rightarrow gg)$ of the SM we can predict $\sigma(pp\rightarrow \tilde\sigma X)$.
The result is compared with the SM Higgs production in Fig.~\ref{fig1}.

The production of $\tilde\sigma$ exceeds that of the 
SM Higgs boson of the same mass $m_{h^0}=m_{\tilde \sigma}$ for $m_{\tilde\sigma}<230$~GeV.
This is because the $\tilde\sigma$ production from $gg$ fusion has an amplitude from the 4-point 
coupling at tree level, while $h^0$ production is governed by the one-loop diagram of the top quark. 
Our prediction of $\sigma(gg\rightarrow \tilde\sigma)$ in Fig.\ref{fig1} 
includes the $\pm 25$\% 
uncertainty associated with the theoretical uncertainty on $\sigma(gg\rightarrow h^0)$.

\noindent\underline{\it Stoponium Decay}\ \ \ \ 
For the $\tilde\sigma$ decay channels $\tilde\sigma\rightarrow AB$,
we consider $AB=gg,\gamma\gamma$,$Z\gamma,W^+W^-$, $ZZ,b\bar b,t\bar t$, and $h^0h^0$.
Their partial widths are given by the formula
\begin{eqnarray}
\Gamma (\tilde\sigma \rightarrow AB) &=& 
\frac{3}{32\pi^2(1+\delta_{AB})} \frac{2p(m_{\tilde\sigma}^2;m_A^2,m_B^2)}{m_{\tilde\sigma}}
\times \frac{|R(0)|^2}{m_{\tilde\sigma}^2}  \sum |{\cal M}|^2\ \ ,
\label{eq2}
\end{eqnarray} 
where ${\cal M}$ represent the free $\tilde t_1\tilde t_1^* \rightarrow AB$ amplitude,
$p$ is the momentum of particle $A$(or $B$) in the CM system, 
and $R(0)$ is the radial wave function of stoponium at the origin.
The total width $\Gamma_{\tilde\sigma}^{\rm tot}$ is the sum of these partial widths. 
Here we omit the LSP neutralino channel $AB=\tilde N_1\tilde N_1$, which is  
a suppressed decay mode\cite{DN,Martin}.  
All the relevant formula are given in the previous works\cite{BK,DN,Martin}, so
we briefly explain here our method and the selection of parameters.

In calculations of the amplitude ${\cal M}={\cal M}(\tilde t_1\tilde t_1^* \rightarrow AB)$, the mass of 
the lighter higgs $h^0$ is fixed to $m_{h^0}=125$~GeV, and a maximal-mixing $\theta_{\tilde t}=\pi/4$
between the two stops $\tilde t_L$ and $\tilde t_R$ is taken.
The value of tan$\beta=v_u/v_d$, the ratio of vacuum expectation values of the two Higgs doublets in SUSY,
is taken to be 10, and the higgs mixing angle $\alpha$ is also fixed from the
tree-level formula, ${\rm tan}2\alpha =[((m_A^2+M_Z^2)/(m_A^2-M_Z^2))$tan$2\beta]$, 
with the choice $m_A=800$~GeV, for which tan$\alpha =-0.103 $ 
.\cite{comment1}

\begin{figure}[htb]
\begin{center}
\resizebox{0.8\textwidth}{!}{
  \includegraphics{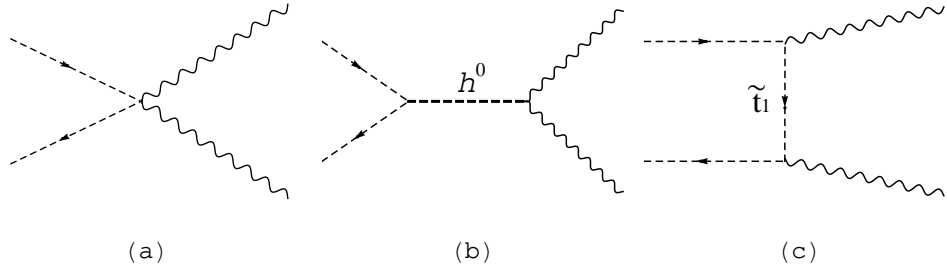}
}
\end{center}
\caption{Dominant generic diagrams of $\tilde\sigma$ decay. Thin dashed lines with arrows represent the
initial $\tilde t_1\tilde t_1^*$ of $\tilde\sigma$ and two wavy lines represent the final states $\bar XX$.
Diagram (a) is taken into account in $\bar XX=gg,\gamma\gamma, Z\gamma$, diagram (b) in $\bar XX=\bar bb$, and
 diagrams (a) and (b) in $\bar XX=WW,ZZ$, while
 all three diagrams contribute to $\bar XX=h^0h^0$.
}
\label{fig2}
\end{figure}

The contribution from the heavier higgs $H^0$ and the heavier stop $\tilde t_2$
 to the amplitudes are neglected under the assumption that they are heavy. 
Amplitudes of $t$-,$u-$ channel $\tilde t$ or $\tilde b$ exchanges are neglected except for the  $\tilde t_1$ exchange
in the $h^0h^0$ channel. Then, all the amplitudes are described by the contact 4-point interaction and/or the
$s$-channel $h^0$ amplitudes when they contribute. 
The Feynman diagrams of our analysis are shown in Fig.~\ref{fig2}.

\begin{figure}[htb]
\begin{center}
\resizebox{0.9\textwidth}{!}{
  \includegraphics{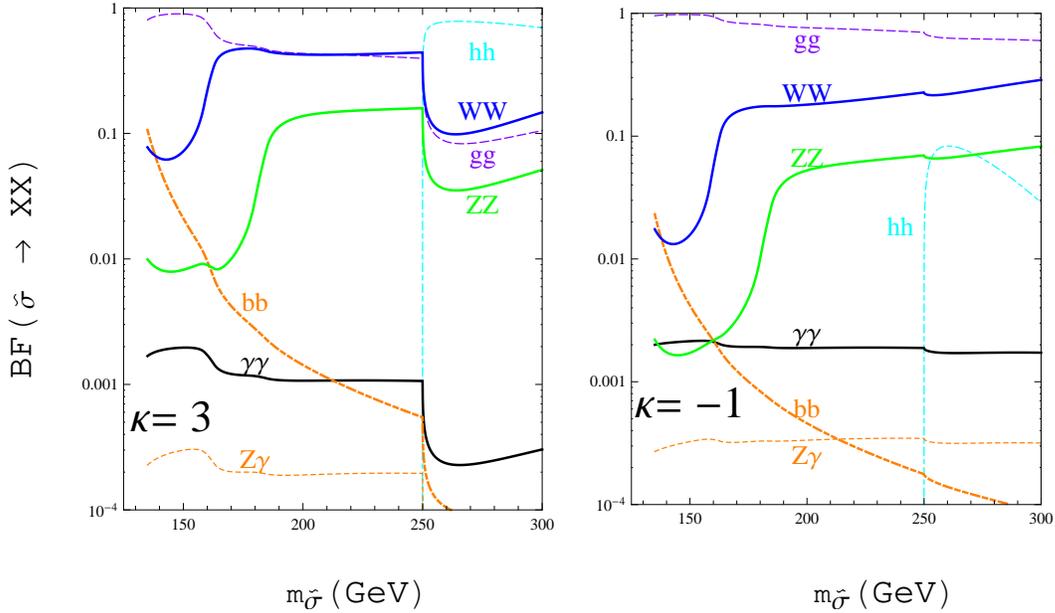}
}
\end{center}
\caption{Decay Branching Fractions of $\tilde\sigma$ versus $m_{\tilde\sigma}$(GeV) 
for $\kappa =3$ and -1. $m_{h^0}$ is taken to be 125 GeV. 
}
\label{fig3}
\end{figure}

For the $gg$ decay of $\tilde\sigma$ 
we include the radiative corrections at NNLO by using the K-factor from 
Higgs production\cite{comment2}.
For decay to $\gamma\gamma$ a special attention is taken by ref.\cite{MY} since this 
branching fraction is larger than that of the Higgs 
 boson for which a strong cancellation between top-quark and W loops occurs.  
We use the $R^{(1)}=\Gamma (\tilde\sigma\rightarrow \gamma\gamma)/
\Gamma (\tilde\sigma\rightarrow{\rm hadrons})$ at NLO in Ref.\cite{MY}, 
equating their
$\Gamma (\tilde\sigma\rightarrow{\rm hadrons})$ to our 
$\Gamma (\tilde\sigma\rightarrow gg)$.
Multiplying $R^{(1)}$ by our $\Gamma(\tilde\sigma\rightarrow gg)$, we  
obtain $\Gamma (\tilde\sigma\rightarrow \gamma\gamma)$ .
The off-shell $WW^*(ZZ^*)$ channels in the low-mass $\tilde\sigma$ case are treated following ref.\cite{Keung}.

The $gg$ partial decay width of $\tilde\sigma$ is proportional\cite{BGHK} to $(|R(0)|/m_{\tilde\sigma})^2$.
We use the non-relativistic quark model with the Wisconsin potential, where 
a potential term in the intermediate range is added to Cornell potential, 
to obtain the value of $|R(0)|$\cite{BGHK,Hagiwara}.    
The binding energy, a few GeV, is much smaller than $m_{\tilde t_1}$ and the stoponium 
mass $m_{\tilde\sigma}$ is well approximated by $2m_{\tilde t_1}$. 

With these simplifications, the results depend only upon two quantities, $m_{\tilde\sigma}$
and the $h^0\tilde t_1\tilde t_1^*$ coupling $\lambda_{h^0\tilde t_1\tilde t_1^*}$, denoted as 
$m_{\tilde t_1}\lambda_2$ in our previous work\cite{BK}. The $\lambda_{h^0\tilde t_1\tilde t_1^*}$ is 
determined by a dimensionless parameter $\kappa$ 
.\cite{comment3}
In our illustrations we consider two values of $\kappa$, $\kappa = 3$ and -1, which correspond to the strong and weak
$h^0\tilde t_1\tilde t_1^*$ coupling, respectively.
The $m_{\tilde\sigma}$ is taken as a free parameter
varying a wide range, 135 GeV $< m_{\tilde\sigma} < 300$~GeV.
Below 135 GeV,
the mixing between $\tilde\sigma$ and $h^0$ (with $m_{h^0}$ taken to be 125~GeV here) may be important,
although this mixing can be handled if necessary.

The decay branching fractions of $\tilde\sigma$ are given in Fig.~\ref{fig3}.
In the case of $\kappa=3$ [larger $\lambda_{h^0\tilde t_1\tilde t_1^*}(=396)$~GeV case],
the branching fractions of $WW,ZZ,h^0h^0,b\bar b,t\bar t$ are larger than those for $\kappa=-1$
[smaller $\lambda_{h^0\tilde t_1\tilde t_1^*}(=169)$~GeV case]
because of larger contributions from the $s$-channel $h^0$ diagram.
$WW, ZZ, h^0h^0$ branching fractions become even larger in the $\kappa =10$ case,
where $WW$ is dominant in $2m_W<m_{\tilde\sigma}<2m_{h^0}$.
In the $\kappa=-1$ case, the decay amplitude to $h^0h^0$ vanishes when $m_{\tilde\sigma}\simeq 370$~GeV
because of a destructive interference among the 4-point interaction, $t$-,$u$-channel $\tilde t_1$ exchange and 
the $s$-channel $h^0$ diagrams. The existence of this cancellation has not been previously noted 
in the literature.

The total width of $\tilde\sigma$ 
in the mass range 
$m_{\tilde\sigma}=135\sim 250$~GeV is fairly small, $\Gamma_{\tilde\sigma}^{\rm tot}\sim 2$ to 7~MeV,
and in the mass range 
$m_{\tilde\sigma}=260\sim 300$~GeV, it is at most $\sim 40$ MeV for the $\kappa=3$ case.
Thus a $\tilde\sigma$ resonance would be observed with the width of the experimental resolution.
At $m_{\tilde\sigma}>2m_W$
the $\Gamma (\tilde\sigma\rightarrow WW)$ partial width is negligibly small compared with 
 $\Gamma (h^0\rightarrow WW)$, and thus $\tilde\sigma$ production via vector boson($WW,ZZ$) fusion
is negligible at the LHC.

The $ZZ$ to $WW$ ratio of the $\tilde\sigma$ decay branching fraction is predicted to be 0.32$\sim$0.36 (for $\kappa=3$) 
in the mass range 200~GeV$ < m_{\tilde\sigma} < 300$~GeV,
as compared with 0.36$\sim$0.44 of the SM $h^0$ in the same mass range.  
This ratio can be used to check if an observed resonance is actually stoponium or not.

\begin{figure}[htb]
\begin{center}
\resizebox{0.8\textwidth}{!}{
  \includegraphics{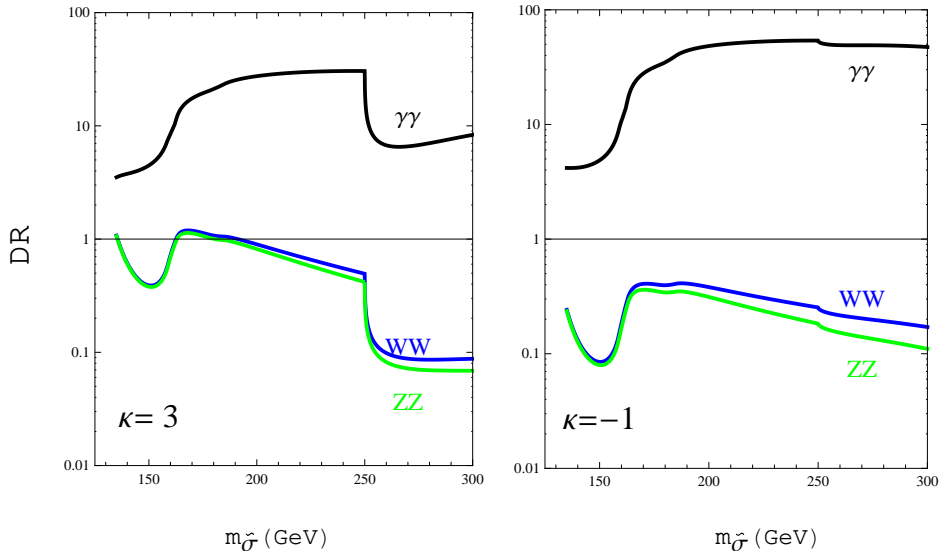}
}
\end{center}
\caption{$\tilde\sigma$ Detection Ratio ($DR$) to the SM higgs $h^0$  of Eq.~(\ref{eq3}) 
for the $\bar XX=W^+W^-$(solid blue), $ZZ$(dashed green), and $\gamma\gamma$(solid black)
final states for $\kappa =3,-1$ versus $m_{\tilde\sigma}$(GeV).
}
\label{fig4}
\end{figure}

\noindent\underline{\it Stoponium Detection compared to SM Higgs}\ \ \ \ \ 
Next we consider the detection of $\tilde\sigma$ in $W^+W^-$, $ZZ$ and $\gamma\gamma$ channels.
The $\tilde\sigma$ search can be made in conjunction with the Higgs search.
The properties of $h^0$  at the LHC are well known, so we use them 
as benchmarks of the search for $\tilde\sigma$.  

The $\tilde\sigma$ detection ratio ($DR$) to $h^0$ in the $\bar XX$ channel is
defined\cite{book} by
\begin{eqnarray}
DR & \equiv &
\frac{\displaystyle \Gamma_{\tilde\sigma\rightarrow gg}\Gamma_{\tilde\sigma\rightarrow \bar XX}/
\Gamma_{\tilde\sigma}^{\rm tot}}{
\displaystyle \Gamma_{h^0\rightarrow gg}\Gamma_{h^0\rightarrow \bar XX}/\Gamma_{h^0}^{\rm tot}}\ ,\ \ \ \ \ \ \ \ \ 
\label{eq3} 
\end{eqnarray}
where $\bar XX=W^+W^-,\ ZZ,$ and $\gamma\gamma$. 
The $DR$ are plotted versus $m_{\tilde\sigma}=m_{h^0}$ in Fig.~\ref{fig4} for the two cases $\kappa=3$ and -1.

In the case of $\kappa =3$ (the large $\lambda_{h^0\tilde t_1\tilde t_1^*}$ case),
the $\tilde\sigma$ to $h^0$ detection ratio is relatively large in both $WW$ and $ZZ$ channels.
The ratio is more than 0.5 in the mass range $160 < m_{\tilde\sigma} < 250$ GeV between the 
WW threshold and the $h^0h^0$ threshold.
For $m_{\tilde\sigma} < 190$ GeV the ratio is nearly unity. 
Even in the $\kappa =-1$ case, the detection ratio is large 
in the $160< m_{\tilde\sigma} < 250$ GeV mass range.

\begin{figure}[htb]
\begin{center}
\resizebox{0.6\textwidth}{!}{
  \includegraphics{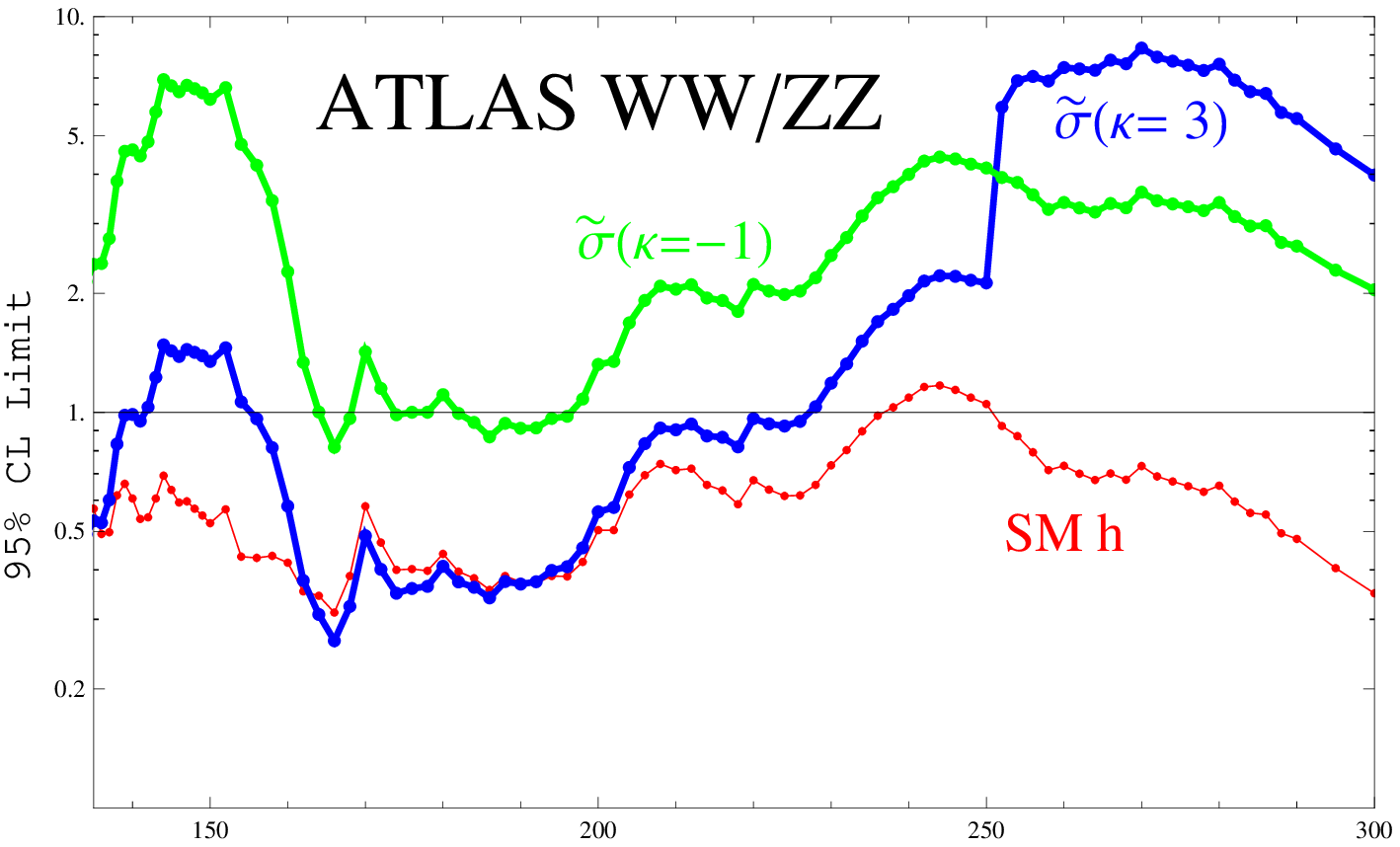}
}
\resizebox{0.6\textwidth}{!}{
  \includegraphics{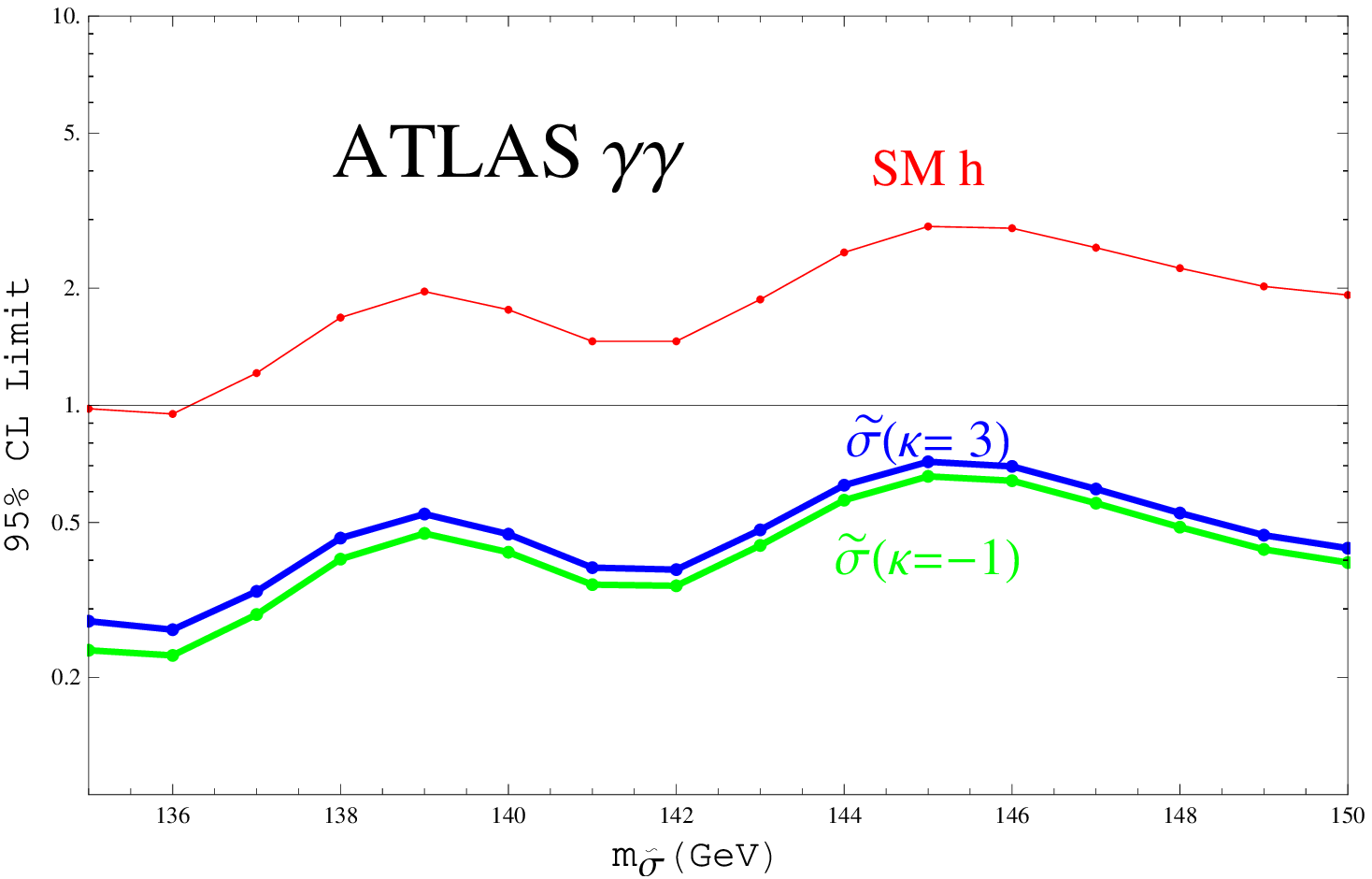}
}
\end{center}
\caption{the 95\% confidence level upper limits of
$(1/DR)\times (\sigma_{\rm exp}/\sigma(gg\rightarrow h^0\rightarrow \bar XX))$.
This is the signal of a boson
decaying into $\bar XX$ relative to the stoponium cross section 
$[\sigma(gg\rightarrow \tilde\sigma \rightarrow \bar XX)=\sigma(gg\rightarrow h^0 \rightarrow \bar XX)\times DR ]$
for $\bar XX=WW$(ATLAS\cite{AtlasWW}) and $\gamma\gamma$
(ATLAS\cite{Atlasll}) data.  
The cases $\kappa =3$(solid blue) and -1(solid green) are shown.
Similar results for the SM Higgs boson are also given(red thin-solid curve).
}
\label{fig5}
\end{figure}

The cross-section of a putative Higgs-boson signal,
relative to the Standard Model cross section, as a function of the assumed Higgs boson mass, 
is widely used by the experimental groups to determine the allowed and excluded regions of $m_{h^0}$.
By use of the $DR$ in Fig.~\ref{fig4}, we can determine the allowed region of
$m_{\tilde\sigma}$ from the present LHC data.
Figure~\ref{fig5} shows the 95\% confidence level upper limits on Higgs-like $\tilde\sigma$ signals
decaying into $\bar XX$ versus $m_{\tilde\sigma}$ for 
$\bar XX=WW$ and $ZZ$ combined(ATLAS\cite{AtlasWW}\cite{comment4}) and 
$\gamma\gamma$(ATLAS\cite{Atlasll}).

For $\kappa=3$, $m_{\tilde\sigma}$ is excluded by ATLAS data over wide ranges of $m_{\tilde\sigma}$
155-227~GeV,
while for the $\kappa =-1$ some regions of $m_{\tilde\sigma}$ are excluded. 
Similar results are found from CMS data\cite{cmsWW}.

The $\tilde\sigma$ search is also applicable to the Tevatron data.
The CDF and D0 experiments excluded the SM Higgs with mass $158\ {\rm GeV}< m_{h^0} < 175$~GeV 
from the data of $WW,ZZ$ channels.
The same data excludes $\tilde\sigma$ in the $\kappa=3$ case in the mass range,
$160\ {\rm GeV} < m_{\tilde\sigma} < 177$~GeV.

The $\gamma\gamma$ final state is very promising for $\tilde\sigma$ detection, because
the $\tilde\sigma$ to $h^0$ detection ratio 
is generally very large in all the mass range of $m_{\tilde\sigma}$, as shown in Fig.~\ref{fig4}.
From the $\gamma\gamma$ data of ATLAS
the region of stoponium mass $135<m_{\tilde\sigma}<150$~GeV is already excluded in both the cases of $\kappa$. 
The $\kappa$-dependence of $DR$ in $\gamma\gamma$
is small below $WW$ threshold because the partial widths of all the allowed two-body final states are independent of $\kappa$.  
For $m_{\tilde\sigma} > 150$~GeV, the $\gamma\gamma$ signal of $h^0$ is too small
to be detected, but the data in this region can determine the existence of $\tilde\sigma$.

\noindent\underline{\it Concluding Remarks}\ \ \ \ 
We have investigated the possibility of finding stoponium $\tilde\sigma$ at LHC7.
In the optimistic case of the stoponium mass and coupling, $\tilde\sigma$ 
 will be discovered in the $WW$, $ZZ$, and $\gamma\gamma$ channels in the search for
the SM Higgs $h^0$. 
The detection rates can be comparable to that of the SM $h^0$,
in the mass region $m_{\tilde\sigma}\sim 160~{\rm GeV}$ up to $2m_{h^0}$ as shown in Fig.~\ref{fig4}. 
The $\gamma\gamma$ search channel is particularly promising since 
$\tilde\sigma$ detection relative to $h^0$ is very large (more than 3) in all mass regions.
A stoponium mass in the 135-150~GeV is already excluded  
in a wide range of supersymmetry parameters from the present ATLAS $\gamma\gamma$ data.  
The $h^0h^0$ decay of $\tilde\sigma$ is another possible mode for discovery, especially for large $\kappa$.

\noindent\underline{\it Acknowledgements}

M.I. is very grateful to the members of phenomenology institute of University of Wisconsin-Madison for hospitalities.
This work was supported in part by the U.S. Department of Energy under grants No. DE-FG02-95ER40896 and
DE-FG02-84ER40173, 
in part by KAKENHI(2274015, Grant-in-Aid for Young Scientists(B)) and in part by grant
as Special Researcher of Meisei University.

\nocite{*}

\bibliography{apssamp}

\begin{thebibliography}{00}
\bibitem{comment0}  In the approximation that the $gg\rightarrow \tilde\sigma$ interaction
 is essentially point-like, the QCD radiative corrections to the tree level $gg\rightarrow h^0$ and $gg\rightarrow \tilde\sigma$
 should be nearly equal so we can use the tree level result for $\Gamma(\tilde\sigma\rightarrow gg)/\Gamma(h^0\rightarrow gg)$.
\bibitem{comment1}
 This formula should be applied with a limit 
 $\alpha \rightarrow \beta -\frac{\pi}{2}$ as $m_A\rightarrow\infty$.
\bibitem{comment2} We use central values of K-factor of Higgs production in NNLO given in Fig. 8 of 
 Ref.\cite{Kfact}: For $m_{h^0}=100\sim 300$~GeV, $K=2.0\sim 2.3$.
 This value is about 10\% larger than the K-factor of Higgs decaying into $gg$ in NNLO given in ref.\cite{decayK} 
 but within the uncertainty of the choice of the renormalization scale. So we simply assume they are equal and adopt the
 value in ref.\cite{Kfact} .
 Here we also note that the $K$ factor for $\tilde\sigma$ production is calculated in NLO\cite{YM} as
 $K\simeq 1.4$ in $m_{\tilde\sigma}=200\sim 600$GeV, which is, contradictorily with our argument\cite{comment0}, 
 25\% smaller than the value 
 $K\simeq 1.9$ in the NLO calculation for $h^0$ given in ref.\cite{Kfact}.
\bibitem{comment3} $\lambda_{h^0\tilde t_1\tilde t_1^*}$ includes a term proportional to 
 $(-\mu\ {\rm sin}\alpha + m_6 A_t\ {\rm cos}\alpha)\equiv \kappa M_W$, where
 $\mu$ is the Higgsino mass term in the superpotential and 
 $m_6 A_t$ is a soft breaking parameter of the trilinear scalar interaction\cite{BK}.
\bibitem{comment4} $DR$ for $WW$ and $ZZ$ are almost the same, and so 
 we have applied $DR$ for $WW$ to the ATLAS $WW$ and $ZZ$ combined data.
\end{thebibliography}

\end{document}